\documentclass[12pt,superscriptaddress,showpacs,showkeys,twocolumn,prb,aps,reprint,a4paper,floatfix]{revtex4-1}
\usepackage{graphicx}
\usepackage{amsmath, amsthm, amssymb, amsfonts}
\usepackage{float}

\usepackage{natbib}
\usepackage{physics}
\usepackage[para,online,flushleft]{threeparttable} 
\usepackage{array}
\usepackage{ragged2e} 

\makeatletter
\newcommand*{\citenst}[2][]{%
  \begingroup
  \let\NAT@mbox=\mbox
  \let\@cite\NAT@citenum
  \let\NAT@space\NAT@spacechar
  \let\NAT@super@kern\relax
  \renewcommand\NAT@open{[}%
  \renewcommand\NAT@close{]}%
  \citet[#1]{#2}%
  \endgroup
}
\makeatother

\makeatletter
\newcommand*{\citenumns}[2][]{%
  \begingroup
  \let\NAT@mbox=\mbox
  \let\@cite\NAT@citenum
  \let\NAT@space\NAT@spacechar
  \let\NAT@super@kern\relax
  \renewcommand\NAT@open{[}
  \renewcommand\NAT@close{]}%
  \cite[#1]{#2}
  \endgroup
}
\makeatother
\usepackage[caption=false]{subfig}

\begin{document}
\title{Calculation and measurement of critical temperature in thin superconducting multilayers}
\author{Songyuan Zhao}
\email{sz311@cam.ac.uk}
\author{D. J. Goldie}
\author{C. N. Thomas}
\author{S. Withington}
\date{\today}

\affiliation{Cavendish Laboratory, JJ Thomson Avenue, Cambridge CB3 OHE, United Kingdom.}

\begin{abstract}
\noindent We have applied the Usadel equations to thin-film multilayer superconductors, and have calculated the critical temperature for general thin-film $S-S'$ bilayer. We extended the bilayer calculation to general thin-film multilayers. The model demonstrates excellent fit with experimental data obtained from Ti-Al bilayers of varying thicknesses.
\end{abstract}


\maketitle

\section{Introduction}
There is increasing interest in proximity-coupled superconducting multilayers. This interest is fuelled by their various potential applications in the fields of transition-edge sensors (TESs) \citenumns{Irwin_1995}, kinetic inductance detectors (KIDs) \citenumns{Day_2003}, superconducting electronics \citenumns{Jarillo-Herrero_2006,Saira_2007,Giazotto_2010}, superconducting transmission line devices \citenumns{Songyuan_transmission_lines_2018}, Josephson junctions \citenumns{Delin_1996,Pepe_2006}, and SIS mixers \citenumns{Dmitiriev_1999}. In the field of TESs, multilayers have been studied to incorporate high conductivity normal metals, and to reproducibly control the transition temperature ($T_c$) \citenumns{Martinis_2000}. In the field of KIDs, multilayers have been studied to achieve tuneable detection frequency thresholds, control over acoustic impedance matching, and protection of vulnerable materials through usage of self-passivating outer layers \citenumns{Catalano_2015,Songyuan_2018,Kaplan_1979,Cardani_2018}.

The superconducting properties of multilayers are governed by their individual layer material properties, geometries, and interface characteristics \citenumns{Brammertz2004}. Calculations of the multilayer $T_c$ from these factors are of considerable value in reducing the time and effort spent on trial-and-error fabrications. In many of the above-mentioned applications, it is important to reliably control the resultant multilayer $T_c$ so as to accommodate experimental needs (for example, bath temperature \citenumns{Songyuan_2018}), and to reproducibly fabricate sensors \citenumns{Martinis_2000}. In addition, such calculations allow the interface characteristics to be determined from the measured $T_c$ \citenumns{Miao_2017_interface, Kushnir_2006}.

The Usadel equations are a set of diffusive-limit equations based on the Bardeen-Cooper-Schrieffer theory of superconductivity. In thick superconducting layers, diffusive-limit equations are applicable in the presence of impurities \citenumns{Anderson_1959}. In thin, clean superconducting films, layer boundaries result in scattering and ensure the applicability of diffusive-limit equations \citenumns{Brammertz2004}. The Usadel equations have been widely used to analyse the $T_c$ of multilayers \citenumns{Zoran_1991,Khusainov_1991,Golubov_1994,Fominov_2001}. In particular, the work by Martinis~{\textit{et al.}} \citenumns{Martinis_2000} provides an analytic analysis of thin-film superconductor-normal conductor ($S-N$) bilayers. The results have been integrated in the analysis and design routines of various multilayer devices, and demonstrate good predictive capabilities \citenumns{Chervenak_2004,Ali_2005,Sadleir_2011,Blois_2013}. More recently, the framework of analysis has been extended to $N-S-N$ trilayers \citenumns{Wang_Tc_2017}.

The Usadel equations have also been numerically solved by Brammertz~{\textit{et al.}} \citenumns{Brammertz_2002} for superconductor-superconductor ($S-S'$) bilayers. The results demonstrate good agreement with measured $T_c$ for Ta-Al and Nb-Al bilayers of various thickness combinations. Whilst a full numerical solution is the most accurate approach to solving the Usadel equations, it is computationally intensive and requires users to be able to implement efficient Usadel equations solvers.

Although the user-friendly results in \citenumns{Martinis_2000} have proven to be extremely useful in the design of bilayer $S-N$ devices, they cannot be applied to $S-S'$ bilayers studied in \citenumns{Zhang_2014, Lolli_2016, Yachin_2017} and general higher-order multilayers studied in \citenumns{Wang_Tc_2017, Cardani_2018, Xiong_2017, Posada_2018}. In view of this, the aim of this paper is to extend the analysis framework of \citenumns{Martinis_2000} to general thin-film multilayer systems.

In section~\ref{sec:Theory} of this paper, we extend the analysis framework of \citenumns{Martinis_2000} to general $S-S'$ bilayer systems. After this, we describe the extension of the bilayer $S-S'$ solution to general trilayer systems and multilayer systems. In section~\ref{sec:Experiment}, we present predictions and measurements of the $T_c$ of Ti-Al bilayer devices of various thickness combinations. The measured $T_c$ data demonstrate good agreement with theoretical predictions, thus giving assurance to the validity of our analysis scheme. We summarize this work in section~\ref{sec:Conclusions}.

\section{Theory}
\label{sec:Theory}
\subsection{Usadel equations and boundary conditions}
\label{sec:Usadel}
The Usadel equations in one dimension are \citenumns{Usadel1970,Brammertz2004,Golubov2004,Vasenko2008}
\begin{equation}
\label{eq:usadel_a}
\frac{\hbar D_S}{2} \pdv[2]{\theta}{x}+iE \sin \theta + \Delta(x) \cos \theta = 0 ,
\end{equation}
and
\begin{equation}
\label{eq:selfCon_a}
\Delta(x)=N_SV_{0,S}\int^{\hbar\omega_{D,S}}_0 dE \operatorname{tanh}\left(\frac{E}{2k_BT}\right)\operatorname{Im}\left(\sin\theta\right) ,
\end{equation}
where $\theta(x,E)$ is a complex variable dependent on position $x$ and energy $E$ parametrising the superconducting properties, $N_S$ is the electron single spin density of states, $V_{0,S}$ is the superconductor interaction potential, $\Delta$ is the superconductor order parameter, $\hbar\omega_{D,S}$ is the Debye energy, $T$ is the temperature of the multilayer, $D_S$ is the diffusivity constant, given by $D_S=\sigma_S/(N_Se^2)$ \citenumns{Martinis_2000}, $e$ is the elementary charge, and finally $\sigma_S$ is the normal state conductivity, at $T$ just above $T_c$. Equation~(\ref{eq:selfCon_a}) is known as the self-consistency equation for order parameter $\Delta(x)$.

The boundary conditions (BCs) relevant to the Usadel equations are presented in \citenumns{Kuprianov1988,Songyuan_2018}.
At the open interface of layer $S$, the BC is given by
\begin{align}
 \pdv{\theta_{S}}{x}=0.
 \label{eq:BC-open}
\end{align}
At the $S'-S$ inner interface, the BCs are given by
\begin{align}
\sigma_{S}\pdv{\theta_{S}}{x}=\sigma_{S'}\pdv{\theta_{S'}}{x},
 \label{eq:BC-inter1}
\end{align}
\begin{align}
R_B\sigma_{S'}\pdv{\theta_{S'}}{x}=\sin(\theta_{S}-\theta_{S'}),
\label{eq:BC-inter2a}
\end{align}
where $R_{B}$ is the product of the boundary resistance between the $S'-S$ layers and its area.
\subsection{Simplifications}
Here we apply the same simplification scheme used in \citenumns{Martinis_2000}:

\begin{enumerate}
  \item At $T$ just above $T_c$, superconductivity is weak, and as a result $|\theta|\ll1$.
  \item The multilayer is thin such that $\theta$ varies slowly, and can be accounted by a second order polynomial expansion.
\end{enumerate}

With these simplifications, the Usadel equations describing a thin multilayer at $T$ just above $T_c$ become
\begin{equation}
\label{eq:usadel}
\frac{\hbar D_S}{2} {\theta''}+ iE \theta + \Delta(x) = 0 ,
\end{equation}
\begin{equation}
\label{eq:selfCon}
\Delta(x)=N_SV_{0,S}\int^{\hbar\omega_{D,S}}_0 dE \operatorname{tanh}\left(\frac{E}{2k_BT_c}\right)\operatorname{Im}\left(\theta\right) ,
\end{equation}
where $\theta''={\partial^2\theta}/{\partial x^2}$.
Equations~(\ref{eq:BC-open},\ref{eq:BC-inter1}) are unchanged. Equation~(\ref{eq:BC-inter2a}) becomes
\begin{align}
\label{eq:BC-inter2}
R_B\sigma_{S'}\theta'_{S'}=\theta_{S}-\theta_{S'}\, ,
\end{align}
where $\theta'={\partial\theta}/{\partial x}$.

Using the Usadel equations and the boundary conditions, one can straightforwardly show that there is a single resulting $T_c$ across a thin-film multilayer: Appendix \ref{sec:One_Tc}.

\subsection{Bilayer Tc calculations}
\label{sec:Bi_theory}
\begin{figure}[ht]
\includegraphics[width=8.6cm]{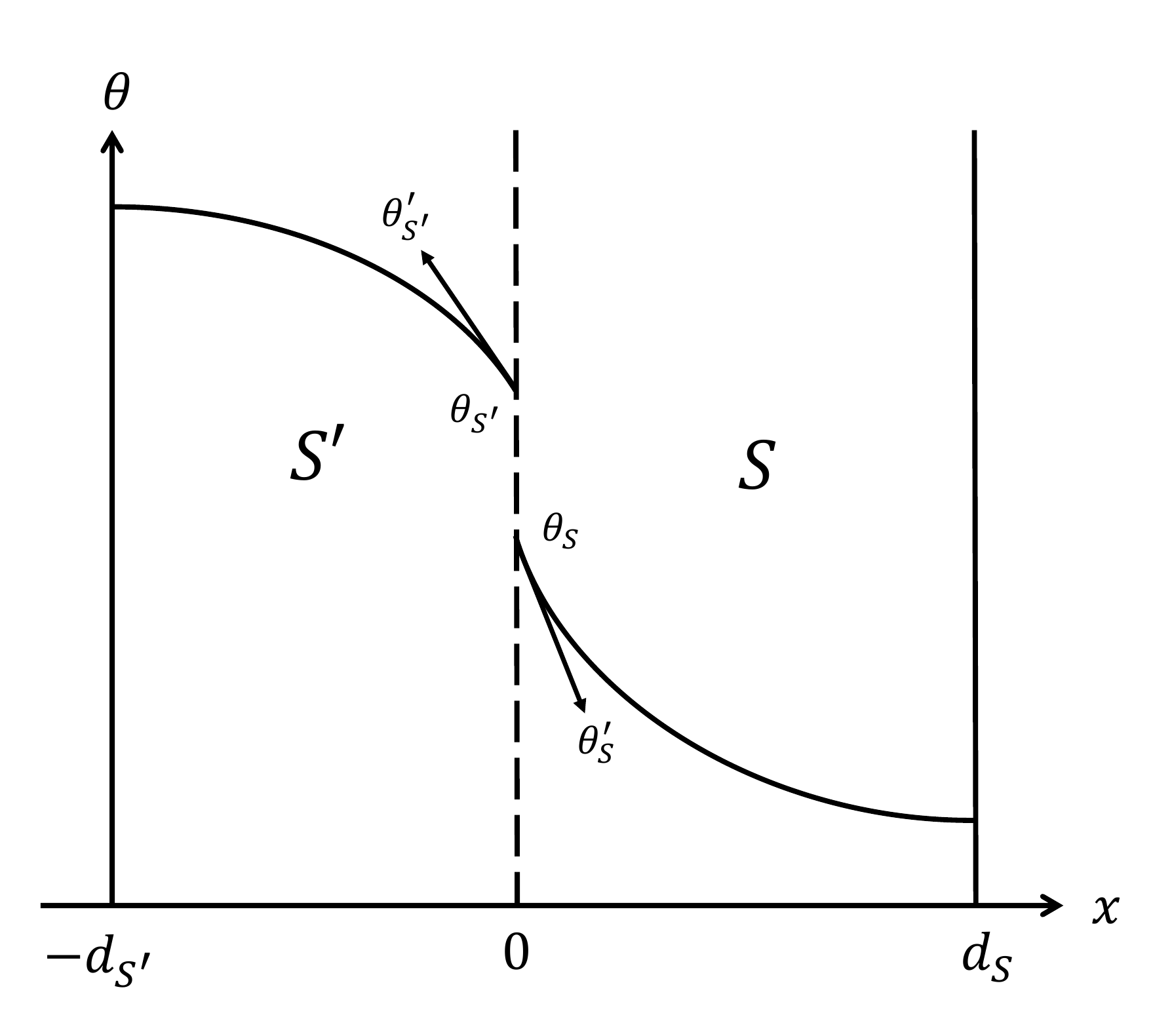}
\caption{\label{fig:Bilayer_geom} Plot of bilayer superconductor parametrization function $\theta$ again position $x$ for a $S'-S$ bilayer.}
\end{figure}
We refer to the geometry shown in figure~\ref{fig:Bilayer_geom}. The interface gradients are given by
\begin{align}
\theta'_{S'} = d_{S'}\theta''_{S'} & = -\frac{2d_{S'}}{\hbar D_{S'}}(\Delta_{S'}+iE\theta_{S'}) \\
\theta'_{S} = -d_{S}\theta''_{S} & = \frac{2d_{S}}{\hbar D_{S}}(\Delta_{S}+iE\theta_{S})
\end{align}
where the gradient is zero at the open boundaries due to equation~(\ref{eq:BC-open}). Formulae for $\theta''_{S'}$ and $\theta''_{S}$ are given by rearrangement of equation~(\ref{eq:usadel}).

The BCs, equations~(\ref{eq:BC-inter1}, \ref{eq:BC-inter2a}), give
\begin{align}
\label{eq:BoundaryM}
-\sigma_{S'}\frac{2d_{S'}}{\hbar D_{S'}}(\Delta_{S'}+iE\theta_{S'})
& = \sigma_{S}\frac{2d_{S}}{\hbar D_{S}}(\Delta_{S}+iE\theta_{S}) \notag \\
& = \frac{1}{R_B}(\theta_{S}-\theta_{S'}) .
\end{align}

Here we introduce a convenient physical constant
\begin{align}
C_{S}=\frac{2d_{S}R_B\sigma_{S}}{\hbar D_{S}}.
\end{align}
Equation~(\ref{eq:BoundaryM}) is used to express $\theta_S$, $\theta_{S'}$ in terms of $\Delta_S$, $\Delta_{S'}$
\begin{align}
\theta_{S}=-\frac{C_{S}\Delta_{S}+C_{S'}\Delta_{S'}-iEC_{S'}C_{S}\Delta_{S}}{iEC_{S}+iEC_{S'}+E^2C_{S}C_{S'}} \\
\theta_{S'}=-\frac{C_{S}\Delta_{S}+C_{S'}\Delta_{S'}-iEC_{S'}C_{S}\Delta_{S'}}{iEC_{S}+iEC_{S'}+E^2C_{S}C_{S'}} .
\end{align}
Taking the imaginary parts
\begin{align}
\operatorname{Im}(\theta_{S})=(f_1C_{S}+f_2)\Delta_{S}+f_1C_{S'}\Delta_{S'} \label{eq:theta_S}\\
\operatorname{Im}(\theta_{S'})=(f_1C_{S'}+f_2)\Delta_{S'}+f_1C_{S}\Delta_{S} \label{eq:theta_S_p}\, ,
\end{align}
where
\begin{align}
f_1=\frac{E(C_{S}+C_{S'})}{(E^2C_{S}C_{S'})^2+E^2(C_{S}+C_{S'})^2}\\
f_2=\frac{E^3C_{S}^2C_{S'}^2}{(E^2C_{S}C_{S'})^2+E^2(C_{S}+C_{S'})^2} \, .
\end{align}
Substituting equations~(\ref{eq:theta_S},\ref{eq:theta_S_p}) into equation~(\ref{eq:selfCon}), we have results of the form
\begin{align}
\Delta_{S}=\left[ A \Delta_{S} + B \Delta_{S'} \right] \\
\Delta_{S'}=\left[ A' \Delta_{S'} + B' \Delta_{S} \right] \, ,
\end{align}
which yields a single equation for $T_c$ that is readily solved numerically
\begin{equation}
\label{eq:Bi_solution}
1=[A+A'-AA'+BB'].
\end{equation}
Here $A,B,A',B'$ are functions of $T_c$, and are given by
\begin{align}
A & =N_{S}V_{0,S}\int_{0}^{\hbar\omega_{D,S}} dE\,\operatorname{tanh}(\frac{E}{2 k_B T_c})\,(f_1C_{S}+f_2) \\
B & = N_{S}V_{0,S}\int_{0}^{\hbar\omega_{D,S}} dE\,\operatorname{tanh}(\frac{E}{2 k_B T_c})\,(f_1C_{S'}) \\
A' & = N_{S'}V_{0,S'}\int_{0}^{\hbar\omega_{D,S'}} dE\,\operatorname{tanh}(\frac{E}{2 k_B T_c})\,(f_1C_{S'}+f_2) \\
B' & = N_{S'}V_{0,S'}\int_{0}^{\hbar\omega_{D,S'}} dE\,\operatorname{tanh}(\frac{E}{2 k_B T_c})\,(f_1C_{S}) .
\end{align}
Practically, $N_{S,S'}$ and $\omega_{D,S,S'}$ are obtained from collated material properties, and $V_{0,S,S'}$ is obtained from the BCS equation \citenumns{Tinkham_1994}
\begin{equation}
k_B T_{c,S,S'} = 1.134\,\hbar\omega_{D,S,S'}\,\operatorname{exp}(\frac{-1}{N_{S,S'}V_{0,S,S'}})
\end{equation}
where $T_{c,S,S'}$ are the measured homogeneous critical temperatures of $S$ and $S'$ layer materials.

Using the above results for $S-S'$ bilayers, and setting the superconductor interaction potential of $S'$ layer $V_{0,S'}=0$, one can recover the results previously derived by Martinis~{\textit{et al.}} \citenumns{Martinis_2000} for $S-N$ bilayers. This confirms our scheme of analysis for general $S-S'$ bilayers in the $S-N$ limit. Analytical results for $S-S'$ bilayers can be obtained in the clean limit where the interface is perfectly transmissive, i.e. $R_B=0$. This result is presented in Appendix \ref{sec:Clean_Tc}.

\subsection{Extension to trilayer and multilayer systems}
In general, there are two types of trilayer systems: 1. The $T_c$ of middle layer is higher / lower than that of both side layers; 2. The $T_c$ of middle layer is between that of the side layers.

\begin{figure}[ht]
\includegraphics[width=8.6cm]{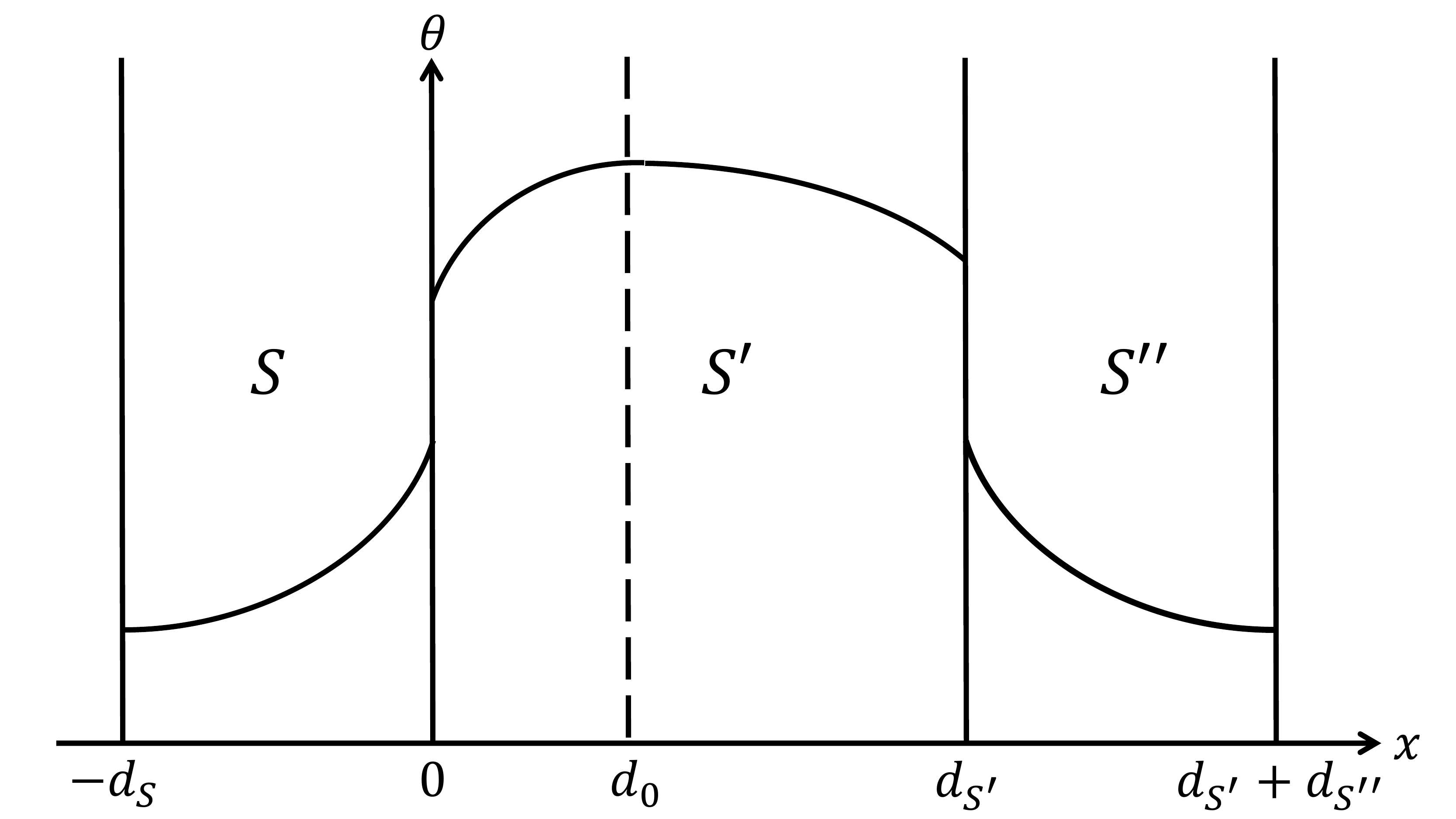}
\caption{\label{fig:Trilayer_case_1} Plot of bilayer superconductor parametrization function $\theta$ again position $x$ for a trilayer, where the middle layer has higher $T_c$ compared to both side layers.}
\end{figure}

For the first case, there exist a $\theta$ maximum/minimum, where $\theta'=0$. Recognizing that $\theta'=0$ is also the open interface boundary condition, we can divide the trilayer into 2 bilayers. As shown in figure~\ref{fig:Trilayer_case_1}, the maximum/minimum location of a $S-S'-S''$ trilayer is denoted as $d_0$. Thus the left $S-S'$ \textit{equivalent} bilayer has thicknesses $d_S$ and $d_0$ respectively, and the right $S'-S''$ \textit{equivalent} bilayer has thicknesses $d_{S'}-d_0$, and $d_{S''}$ respectively. Equation~(\ref{eq:Bi_solution}) in the previous section attains the form
\begin{align}
  F_L(d_0,T_c)&=1 \, , \, \mathrm{and} \label{eq:Tri_first}\\
  F_R(d_0,T_c)&=1 \, , \label{eq:Tri_second}
\end{align}
where $F_L(d_0,T_c)=1$ is the analogue of equation~(\ref{eq:Bi_solution}) in the left $S-S'$ \textit{equivalent} bilayer, $F_R(d_0,T_c)=1$ is the analogue of equation~(\ref{eq:Bi_solution}) in the right $S'-S''$ \textit{equivalent} bilayer, and each has solution that traces out a curve of $T_c$ as a function of $d_0$. The intersection between the two $T_c(d_0)$ curves gives the solution for the overall device $T_c$. In practice, the coupled pair of equations~(\ref{eq:Tri_first},\ref{eq:Tri_second}) can be solved using a standard numerical solver.

The virtue of the above analysis is that it can be straightforwardly extended to a general N-layer system where the maxima/minima of $\theta$ can be found in all middle layers. For a general N-layer system, there are N-2 middle layers and as a result N-1 equivalent bilayers. Denote $d_i$ as the extremal point of the i-th middle bilayer. There are thus N-1 unknowns (N-2 values of $d_i$ and $T_c$), and N-1 analogues of equation~(\ref{eq:Bi_solution}). This system can also be solved using a standard numerical solver.

\begin{figure}[ht]
\includegraphics[width=8.6cm]{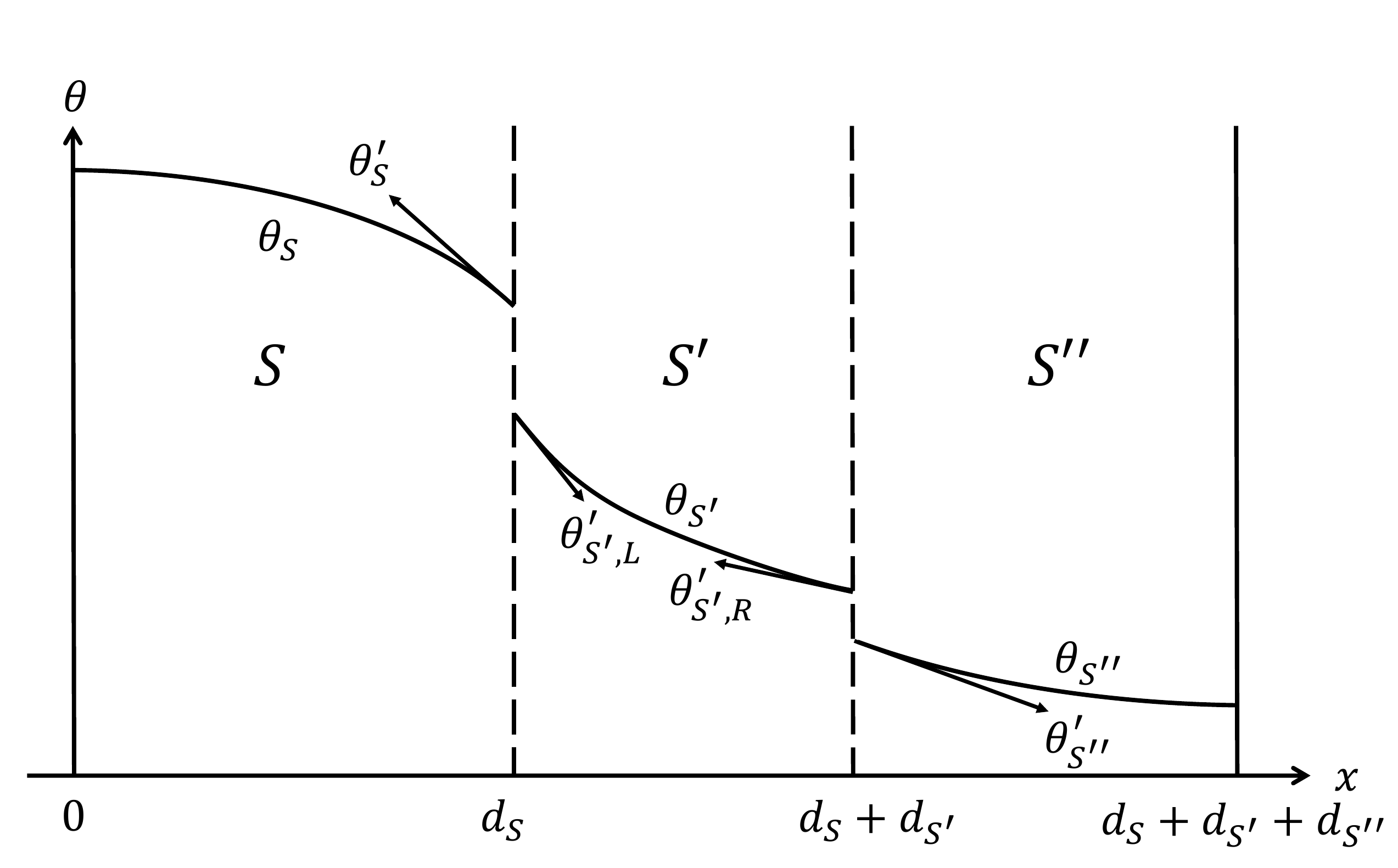}
\caption{\label{fig:Trilayer_case_2} Plot of bilayer superconductor parametrization function $\theta$ again position $x$ for a trilayer, where the middle layer has $T_c$ that is in between both side layers.}
\end{figure}

For the second case of a $S-S'-S''$ trilayer without maximum/minimum in the middle layer, as shown in figure~\ref{fig:Trilayer_case_2}, we need to connect the left and right inner boundaries via
\begin{align}
\theta'_{S',R} &= \theta'_{S',L} + d_{S'}\theta''_{S'} \\
&= \theta'_{S',L} -\frac{2d_{S'}}{\hbar D_{S'}}(\Delta_{S'}+iE\theta_{S'}) \, .
\end{align}
Using the same analysis method in section~\ref{sec:Bi_theory}, $\theta''$ above is substituted away using equation~(\ref{eq:usadel}). Again noting that at the open boundaries, $\theta'=0$, it can then be shown that
\begin{align}
\theta'_{S',L} &= -\frac{\sigma_{S}}{\sigma_{S'}}\frac{2d_{S}}{\hbar D_{S}}(\Delta_{S}+iE\theta_{S}) \\
&= \frac{1}{\sigma_{S'}R_B}(\theta_{S'}-\theta_{S})\\
\theta'_{S',R} &= \frac{\sigma_{S''}}{\sigma_{S'}}\frac{2d_{S''}}{\hbar D_{S''}}(\Delta_{S''}+iE\theta_{S''}) \\
&= \frac{1}{\sigma_{S'}R_B}(\theta_{S''}-\theta_{S'}) \, .
\end{align}

From the above set of equation, $\theta_{S,S',S''}$ can be expressed in terms of $\Delta_{S,S',S''}$ through simple rearrangements. These can then be substituted into equation~(\ref{eq:selfCon}), linking $\Delta_{S,S',S''}$ with each other. After these manipulations, we now have three unknowns ($\Delta_{S}$, $\Delta_{S'}$, and $\Delta_{S''}$) and three equations. Through some algebraic cancelling, a final equation of the form $F_{\mathrm{tri}}(T_c)=1$, akin to equation~(\ref{eq:Bi_solution}), can again be obtained.

The same procedure can be employed for higher-order multilayer systems. For each additional middle layer $S'$, five new unknowns are introduced ($\theta_{S'}$, $\theta'_{S',L}$, $\theta'_{S',R}$, $\theta''_{S'}$ and $\Delta_{S'}$). These five unknowns are compensated by one more analogue of equation~(\ref{eq:usadel}), one more geometry equation linking $\theta'_{S',L}$ and $\theta'_{S',R}$, one more self-consistency equation linking $\theta_{S'}$ and $\Delta_{S'}$, and two more boundary conditions.

\section{Experimental Results}
\label{sec:Experiment}
\subsection{Fabrication Details}
Films were deposited onto $50\,\,\rm{mm}$ diameter Si wafers by DC magnetron sputtering at a base pressure of $2\times 10^{-10}\,\,\rm{Torr}$  or below. Ti films were deposited at ambient temperature and Al films were deposited after substrate cooling to liquid nitrogen temperatures. Bilayer Ti-Al films were deposited without breaking vacuum. Titanium is deposited first for all bilayers studied. The wafers were diced into $13.5\times7.5\,\,\rm{mm}$ samples and electrical contacts were made to the unpatterned films in a 4-wire geometry using Al wirebonds. The samples were attached to the cold stage of a dilution refrigerator and the temperature was monitored using a calibrated ruthenium oxide thermometer. Resistance measurements were made with an AC resistance bridge with bias currents of $3\,\, \textrm{or}\,\, 30\,\,\rm{\mu A}$.

\subsection{$T_c$ Measurements}
\begin{table}[ht]
\begin{threeparttable}
\caption{\label{tab:table1}Table of material properties.}
\begin{tabular}{b{0.40\linewidth} b{0.26\linewidth} b{0.28\linewidth}}
\toprule
 & \textrm{Aluminium} & \textrm{Titanium} \\
\colrule
$T_{\mathrm {c}}$ (K) & 1.20\tnote{a} & 0.550\tnote{b}, 0.588\tnote{c} \\
$\sigma_{\mathrm{N}}$ (/$\mu\Omega\,$m) \tnote{d} & 132\tnote{a} & 5.88\tnote{a} \\
$RRR$ \tnote{e} & 5.5\tnote{a} &  3.5\tnote{a} \\
$N_0$ ($10^{47}$/$\textrm{J}\,\textrm{m}^3$) & 1.45\tnote{f} & 1.56\tnote{f} \\
$D$ ($\mathrm{m^2s^{-1}}$) & 35\tnote{g} &  1.5\tnote{g} \\
$\xi$ (nm) & 189\tnote{h} & 57\tnote{h} \\
$\Theta_D$ (K) & 423\tnote{i} & 426\tnote{i} \\
\toprule
\end{tabular}
\begin{tablenotes}[flushleft]
\RaggedRight
\footnotesize
\item[a] Measured. \\
\item[b] Measured for the first set of measurements in December 2017. \\
\item[c] Measured for the second set of measurements in May 2018. \\
\item[d] $\sigma_{\mathrm{N}}$ is the normal state conductivity. \\
\item[e] $RRR$ is the residual resistivity ratio.\\
\item[f] $N_0$ is the normal state electron density of states, and is calculated from the free electron model \citenumns{Ashcroft_1976}. \\
\item[g] Diffusivity constant $D$ is calculated using $D_s = \sigma_{N,s}/(N_{0,s}e^2)$ \citenumns{Martinis_2000}. \\
\item[h] Coherence length $\xi$ is calculated using $\xi_s=[{\hbar D_s/(2\pi k_\mathrm {B} T_\mathrm {c}})]^{1/2}$ \citenumns{Brammertz2004}, where $k_\mathrm {B}$ is the Boltzmann constant. \\
\item[i] $\Theta_D$ is the Debye temperature, and is given by $k_B\Theta_D=\hbar \omega_D$. Values are taken from \citenumns{Gladstone}.
\end{tablenotes}
\end{threeparttable}
\end{table}

\begin{figure}[!ht]
\includegraphics[width=8.6cm]{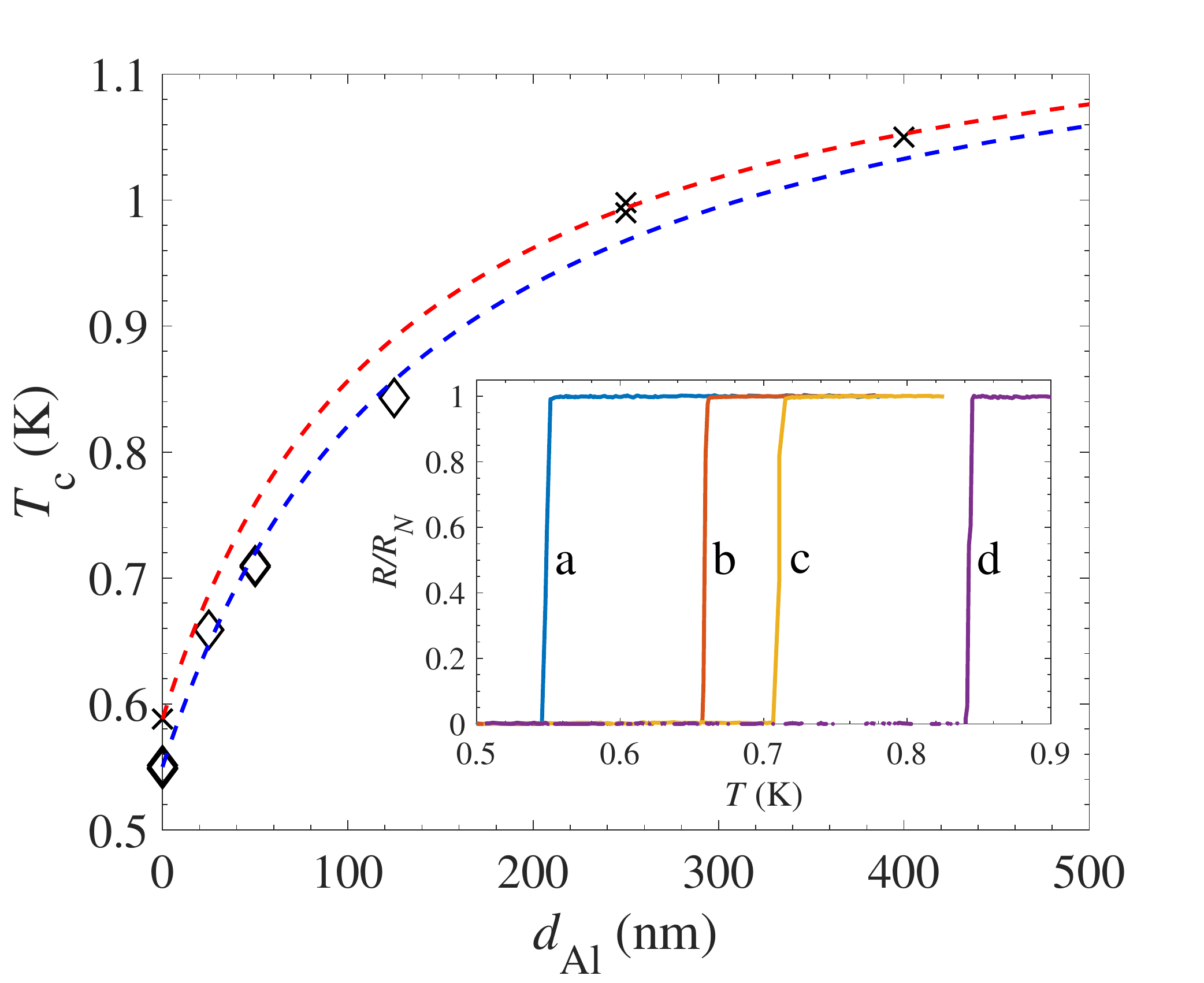}
\caption{\label{fig:expt_fit} Plot of superconducting transition temperature $T_c$ versus Al thickness $d_{\mathrm{Al}}$, for fixed Ti thickness $d_{\mathrm{Ti}}=100\,\textrm{nm}$. Diamonds indicate Ti-Al bilayers measured in December 2017. Crosses indicate Ti-Al bilayers measured in May 2018. The dashed, blue line is the theory plot for diamonds, and is calculated with $R_B=(0^{+1}_{-0})\times 10^{-16}\,\Omega\mathrm{m^2}$, $T_{c,\textrm{Ti}}=0.55\,\textrm{K}$. The dashed, red line is the theory plot for crosses, and is calculated with $R_B=(0^{+4}_{-0})\times 10^{-17}\,\Omega\mathrm{m^2}$, $T_{c,\textrm{Ti}}=0.588\,\textrm{K}$. Inset: normalized resistance $R/R_N$ versus temperature for solid lines (a) $d_{\mathrm{Al}}=0\,\textrm{nm}$ Ti-Al bilayer, (b) $d_{\mathrm{Al}}=25\,\textrm{nm}$ Ti-Al bilayer, (c) $d_{\mathrm{Al}}=50\,\textrm{nm}$ Ti-Al bilayer, and (d) $d_{\mathrm{Al}}=125\,\textrm{nm}$ Ti-Al bilayer. Here $R$ is measured resistance, and $R_N$ is measured the normal state resistance.}
\end{figure}

Two sets of measurements were taken. The first set of measurements was performed in late 2017 for 7 Ti-Al bilayers with $d_{\mathrm{Al}}$ ranging from $0\,\textrm{nm} - 125\,\textrm{nm}$. The second set of measurements was performed 6 months later for 4 Ti-Al bilayers with $d_{\mathrm{Al}}$ ranging from $0\,\textrm{nm} - 400\,\textrm{nm}$. $d_{\mathrm{Ti}}$ is fixed at $100\,\textrm{nm}$ for both sets of measurements. To account for slight variations in the sputtering system, we measure $T_{c,\textrm{Ti}}$ for each set of measurements. The value of $T_{c,\textrm{Ti}}$ from the second set of measurements is $7\%$ higher than that from the first.

The results of bilayer $T_c$ measurements are shown in figure~\ref{fig:expt_fit}, plotted against $d_{\mathrm{Al}}$. Diamonds indicate Ti-Al bilayers measured in December 2017. Crosses indicate Ti-Al bilayers measured in May 2018. The dashed, blue line is the theory plot for diamonds, and is calculated with $R_B=(0^{+1}_{-0})\times 10^{-16}\,\Omega\mathrm{m^2}$, $T_{c,\textrm{Ti}}=0.55\,\textrm{K}$. The best-fit value of $R_B$ is obtained by minimizing the squared-error between the model and the data, and the uncertainty indicates the value of $R_B$ where the squared-error increases by a factor of $1/e$, where $e$ is Euler's number. The dashed, red line is the theory plot for crosses, and is calculated with $R_B=(0^{+4}_{-0})\times 10^{-17}\,\Omega\mathrm{m^2}$, $T_{c,\textrm{Ti}}=0.588\,\textrm{K}$. The theory plots are generated using the physical parameters listed in Table~\ref{tab:table1}. The values of $R_B$ are small for both theory plots, and indicate that our deposition technique achieves very clean layer interfaces consistently. Given that the deposited bilayer interfaces are consistently very clean ($R_B \approx 0\,\Omega\mathrm{m^2}$), and that we have measured both $T_{c,\textrm{Ti}}$ and $T_{c,\textrm{Al}}$ for each set of measurements, our theory plots are effectively generated without free parameter. The measured $T_c$ values of Ti-Al bilayers demonstrate good agreement with the theoretical plot, with $\chi^2=1\times10^{-3}$ for the December 2017 dataset, and $\chi^2=4\times10^{-5}$ for the May 2018 dataset. This close agreement between theory and measurements lends confidence to the validity of our analysis scheme. Multiple measurements of three $d_{\mathrm{Al}}=0\,\textrm{nm}$ bilayers (diamond data points) overlay closely with each other, demonstrating the reproducibility of our depositions. Similar closeness of measured $T_c$ is observed in the two $d_{\mathrm{Al}}=50\,\textrm{nm}$ bilayers.
The inset shows the normalized resistance $R/R_N$ versus temperature for solid lines (a) $d_{\mathrm{Al}}=0\,\textrm{nm}$ Ti-Al bilayer, (b) $d_{\mathrm{Al}}=25\,\textrm{nm}$ Ti-Al bilayer, (c) $d_{\mathrm{Al}}=50\,\textrm{nm}$ Ti-Al bilayer, and (d) $d_{\mathrm{Al}}=125\,\textrm{nm}$ Ti-Al bilayer. $R$ is the measured resistance, and $R_N$ is the normal state resistance just above $T_c$. These bilayers display sharp superconducting state transitions, typically having transition widths $\Delta T\sim3\,\textrm{mK}$.

\section{Conclusions}
\label{sec:Conclusions}
In this paper, we have described a general analysis of thin-film multilayer $T_c$, using the diffusive-limit Usadel equations. We performed a derivation of $T_c$ for a general thin-film $S-S'$ bilayer. We have described methods of extending this calculation to general thin-film trilayers and multilayers. Our model for $S-S'$ bilayer reduces to previous results in \citenumns{Martinis_2000} when the $T_c$ of $S'$ layer is set to zero. Our model is easy to implement and computationally fast. We find a five order-of-magnitude reduction in $T_c$ calculation time compared to solving the full Usadel equations as presented in \citenumns{Songyuan_2018}. Our experimental measurements of Ti-Al bilayer $T_c$ demonstrate good agreement with predictions from our model, thereby enabling our analysis method to be incorporated in the design of superconducting multilayer devices. Work is currently being conducted to extract $R_B$ of multilayers with less transmissive boundaries using our model, and is proving successful.

\appendix
\section{Critical temperature across layers}
\label{sec:One_Tc}
 We consider a $S-S'$ bilayer whilst relaxing the assumption of a single \textit{resulting} $T_c$. We define $T_{c,1}$ and $T_{c,2}$ to be the \textit{resulting} critical temperature of layer $S$ and $S'$ respectively. Without loss of generality, we can order the layer notation such that $T_{c,1}<T_{c,2}$.

In layer $S$, at $T>T_{c,1}$, by definition of the critical temperature,
\begin{align}
  \Delta_{S}&=0\\
  \theta_{S}&=0 \, .
\end{align}
Using equation~(\ref{eq:usadel}), we deduce $\theta''_{S}=0$. In the case of thin films, we have $\theta'_{S}=d_{S}\theta''_{S}=0$. In layer $S'$, BC equation~(\ref{eq:BC-inter1}) ensures that $\theta'_{S'}=0$, and equation~(\ref{eq:BC-inter2}) ensures that $\theta_{S'}=0$.

Thus if layer $S$ is in the normal state, layer $S'$ is also in the normal state, i.e. if $T>T_{c,1}$, then $T>T_{c,2}$. From the ordering of the layers, $T_{c,1}<T_{c,2}$, we thus conclude that $T_{c,1}=T_{c,2}=T_c$, i.e. there is a single $T_c$ across the entire device. The above argument can be extended to a general thin-film multilayer to show that there is a single $T_c$ across the entire multilayer. This can be done by repeatedly applying equation~(\ref{eq:BC-inter1}, \ref{eq:BC-inter2}) across each layer interface.

In this case of metamaterials made up of thin periodic multilayers, this phenomenon of a single resulting $T_c$ has been observed experimentally in \citenumns{Smolyaninova_2016}.

\section{Analytical bilayer solution in the clean limit}
\label{sec:Clean_Tc}
The clean interface limit is when $R_B = 0$. Using equation~(\ref{eq:BC-inter2}), the clean limit implies that $\theta_{S}=\theta_{S'}$.
Here we introduce another convenient constant
\begin{align}
G_{S}=\frac{2d_{S}\sigma_{S}}{\hbar D_{S}}.
\end{align}
Equation~(\ref{eq:BoundaryM}) is used to express $\theta_S$, $\theta_{S'}$ in terms of $\Delta_S$, $\Delta_{S'}$
\begin{align}
\theta_{S}=i\operatorname{Im}(\theta_{S})&=\theta_{S'}=i\operatorname{Im}(\theta_{S'}) \\
&=i\frac{1}{E}\frac{G_S\Delta_S+G_{S'}\Delta_{S'}}{G_S+G_{S'}}.
\end{align}
Substituting the above equations into equation~(\ref{eq:selfCon}), we again have results of the form
\begin{align}
\Delta_{S}=\left[ A \Delta_{S} + B \Delta_{S'} \right] \label{eq:clean_system}\\
\Delta_{S'}=\left[ A' \Delta_{S'} + B' \Delta_{S} \right] \, . \label{eq:clean_system_2}
\end{align}
We can simplify coefficients $A,B,A',B'$ immensely due to the simple $1/E$ dependence of $\operatorname{Im}(\theta_{S})$ and $\operatorname{Im}(\theta_{S'})$. We note that
\begin{align}
  \int_{0}^{\hbar\omega_{D}} dE\,\frac{1}{E}\operatorname{tanh}(\frac{E}{2 k_B T_c}) &= \int_{0}^{\frac{\Theta_D}{2T_c}} dx\,\frac{1}{x}\operatorname{tanh}(x) \\
  & \approx \operatorname{ln}\left(\frac{2e^\gamma}{\pi}\frac{\Theta_D}{T_c}\right) \, ,
\end{align}
where $\gamma\approx0.57721$ is the Euler–Mascheroni constant. The logarithmic approximation to the integral is valid in the regime that ${\Theta_D}/{(2T_c)}\gg1$.

We define another convenient constant $K = \frac{2e^\gamma}{\pi}$. Using the BCS result
\begin{align}
  \frac{1}{N_{S}V_{0,S}} &= \int_{0}^{\hbar\omega_{D,S}} dE\,\frac{1}{E}\operatorname{tanh}(\frac{E}{2 k_B T_{c,S}}) \\
   &=\operatorname{ln}\left(K\frac{\Theta_{D,S}}{T_{c,S}}\right)\, ,
\end{align}
we can explicitly find the coefficients
\begin{align}
 A &= \frac{G_S}{G_S+G_{S'}}\operatorname{ln}\left(K\frac{\Theta_{D,S}}{T_c}\right)/\operatorname{ln}\left(K\frac{\Theta_{D,S}}{T_{c,S}}\right) \\
 B &= \frac{G_{S'}}{G_S+G_{S'}}\operatorname{ln}\left(K\frac{\Theta_{D,S}}{T_c}\right)/\operatorname{ln}\left(K\frac{\Theta_{D,S}}{T_{c,S}}\right)\\
 A' &= \frac{G_{S'}}{G_S+G_{S'}}\operatorname{ln}\left(K\frac{\Theta_{D,S'}}{T_c}\right)/\operatorname{ln}\left(K\frac{\Theta_{D,S'}}{T_{c,S'}}\right)\\
 B' &= \frac{G_{S}}{G_S+G_{S'}}\operatorname{ln}\left(K\frac{\Theta_{D,S'}}{T_c}\right)/\operatorname{ln}\left(K\frac{\Theta_{D,S'}}{T_{c,S'}}\right)\,.
\end{align}
By substituting the above coefficients into equations~(\ref{eq:clean_system},\ref{eq:clean_system_2}), the following formula for $T_c$ can be readily obtained
\begin{align}
  T_c = \exp{\frac{H_{S}\operatorname{ln}\left(K\Theta_{D,S}\right)+H_{S'}\operatorname{ln}\left(K\Theta_{D,S'}\right)-1}{H_{S}+H_{S'}}} \, , \label{eq:clean_Tc}
\end{align}
where
\begin{align}
  H_{S} = {\frac{G_S}{G_S+G_{S'}}}/{\operatorname{ln}\left(K\frac{\Theta_{D,S}}{T_{c,S}}\right)}\\
  H_{S'} = {\frac{G_{S'}}{G_S+G_{S'}}}/{\operatorname{ln}\left(K\frac{\Theta_{D,S'}}{T_{c,S'}}\right)}\, .
\end{align}
\bibliographystyle{IOP_no_URL}
\bibliography{library}
\end{document}